\documentclass{ws-ijmpcs}

\begin{document}

\markboth{J. Hawkes et al.}
{Investigation of Dense Gas Towards Relativistic Outflow Sources }

%
\catchline{}{}{}{}{}
%

\title{INVESTIGATION OF DENSE GAS TOWARDS RELATIVISTIC OUTFLOW SOURCES}

\author{J. HAWKES$^{\ast}$,  G. ROWELL$^{\ast}$, B. DAWSON$^{\ast}$, F. AHARONIAN$^{\dag}$, M. BURTON$^{\ddag}$, Y. FUKUI$^{\S}$, N. FURUKAWA$^{\S}$, T. HAYAKAWA$^{\S}$, A. KAWAMURA$^{\S}$, N. MAXTED$^{\ast}$, E. de O\~{N}A-WILHELMI$^{\dag}$, P. de WILT$^{\ast}$ and A. WALSH$^{\P}$}
\address{$^{\dagger}$School of Chemistry and Physics, University of Adelaide, SA, 5005, Australia,  $^{\dag}$Max-Planck Institut f\"{u}r Kernphysik, P.O. Box 103980, D-69029 Heidelberg, Germany, $^{\ddag}$School of Physics, University of New South Wales, Sydney 2052, Australia, $^{\S}$Dept. Astrophysics, Nagoya University, Chikusa-ku, Nagoya, 464-8062, Japan, $^{\P}$International Centre for Radio Astronomy Research, Curtin University, GPO Box U1987, Perth, Australia}

\maketitle

\begin{abstract}
We probe the interstellar medium towards the objects Circinus X-1, a low-mass X-ray binary with relativistic jets; and the highly energetic Westerlund 2 stellar cluster, which is located towards TeV gamma-ray emission and interesting arc- and jet-like features seen in Nanten $^{12}$CO data. We have mapped both regions with the Mopra radio telescope, in 7 mm and 12 mm wavebands, looking for evidence of disrupted/dense gas caused by the interaction between high energy outflows and the ISM. Towards Westerlund 2, peaks in CS(J=1-0) emission indicate high density gas towards the middle of the arc and the endpoint of the jet; and radio recombination line emission is seen overlapping the coincident HII region RCW49. Towards Circinus X-1, $^{12}$CO(J=1-0) Nanten data reveals three molecular clouds that lie in the region of Cir X-1. Gas parameters for each cloud are presented here.

\end{abstract}

\section{Introduction}	
Circinus X-1 (Cir X-1) is a low-mass X-ray binary and microquasar, containing a young neutron star with an orbital period about its companion of $\sim$16.6 days\cite{Tudose}. Its highly relativistic jets (apparent velocity $\beta_{app} >$ 12)\cite{Tudose} are resolved in both radio and X-ray wavelengths\cite{Stewart,Haynes} and interact with the interstellar medium (ISM) to create arcminute-scale synchrotron X-ray lobes\cite{Sell} and a 5$'$ diameter radio nebula surrounding Cir X-1\cite{Tudose}. Given the evidence of multi-TeV particle acceleration in the region, Cir X-1 is a good candidate for TeV gamma-ray ($\gamma$-ray) emission, however no TeV $\gamma$-ray emission has been seen with H.E.S.S. (28 hrs)\cite{Nicholas}. Our observations of the molecular gas towards Cir X-1 aim to provide insights into the physics of jet/ISM interactions on parsec (pc) scales, assess the potential of Cir X-1 as a cosmic-ray accelerator and provide additional constraints on its distance, which is ambiguous at 4-12 kpc.

Westerlund 2 is a young (1-3 Myr) and rich stellar cluster containing ongoing star formation and over two dozen high mass stars, including two Wolf-Rayet stars; and ionizes the nearby giant HII region RCW 49. Interesting arc- and jet-like features seen in Nanten $^{12}$CO data\cite{Fukui09} towards the cluster may be an indication of relativistic particle acceleration in the region and may have a common origin with coincident TeV $\gamma$-ray source HESS J1023-575\cite{Aharonian}.  Given the uncertainty in the origin of the TeV source and its link to the cluster, arc-jet feature and nearby pulsar PSR J1022-5746, a detailed picture of the dense and disrupted gas in this region is needed. 

\section{Observations}
The area towards Cir X-1 was mapped in the $^{12}$CO(J=1-0) line as part of the Nanten Galactic Plane Survey\cite{Mizuno}. Follow-up observations with the Mopra radio telescope targeted tracers of denser and disrupted gas and were conducted in February 2012 in the 7 mm band and January 2010, February 2010 and April 2011 in the 12 mm band with a resultant T$_{RMS}$ of 0.1 K and 0.07 K respectively. Towards Westerlund 2, observations in the 12 mm band were conducted in January 2012 and were comprised of three 30$'\times$30$'$ and one 10$'\times$10$'$ mapping region to result in full coverage of the HII region, TeV emission and molecular arc-jet feature. The 7 mm observations were conducted in April 2012 and had coverage as seen in Fig.~\ref{f2}. The T$_{RMS}$ was $\sim$0.08 K per channel in all cases. The observations with the Mopra radio telescope in the 7 and 12 mm bands and data reduction were conducted as in Ref.~\refcite{Maxted}.

\section{Results}
\begin{figure}[!htb]
\centerline{\psfig{file=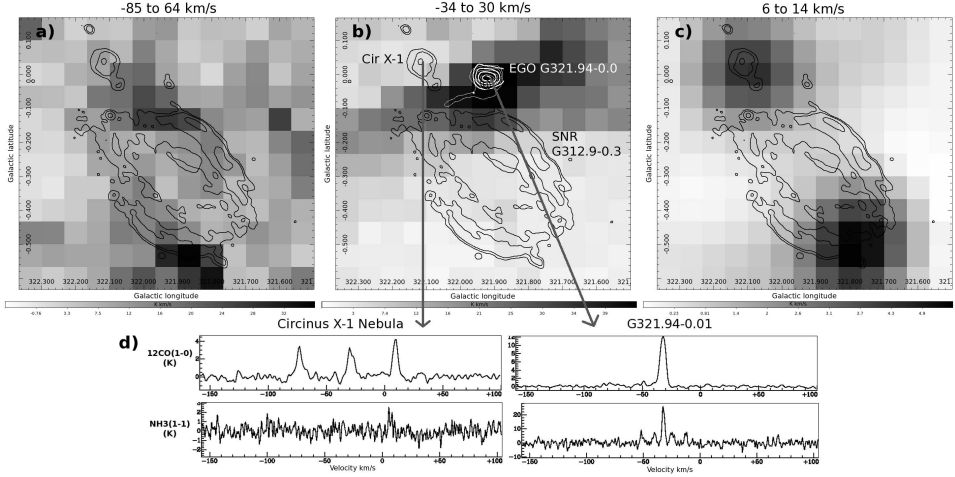,width=12.7cm}}
\vspace*{8pt}
\caption{Cir X-1 region, showing velocity integrated Nanten $^{12}$CO(J=1-0) data in units of K km/s, overlaid with MOST 843 MHz radio continuum contours (black) that show the radio nebula and nearby SNR G312.9-0.3, which is not thought to be associated with Cir X-1\protect\cite{Bhattacharya}.  
a)$-$85 to $-$64 km/s. The cloud has kinematic distance 4 or 11 kpc and the molecular mass inside an ellipse of 109 arcsec$^{2}$ encompassing the SNR and flanking CO emission is $\sim$2x10$^{5}$ M$_{\odot}$ (d/5.8kpc)$^{2}$. 
b)$-$34 to $-$30 km/s with additional CS(J=1-0) contours (thin white lines) in steps of 1$\sigma$ from 2$\sigma$ and NH$_{3}$(1,1) contours (thick white lines) in steps of 3$\sigma$ from 5$\sigma$. The cloud has kinematic distance 2 or 13 kpc and the total molecular mass in the cloud is $\sim$4x10$^{5}$ M$_{\odot}$ (d/10kpc)$^{2}$. 
c)6 to 14 km/s. The northern cloud has kinematic distance $<$ 16 kpc and has mass of a few x10$^{4}$ M$_{\odot}$ (d/10kpc)$^{2}$ and density $\sim$70 cm$^{-3}$ (d/10kpc)$^{-1}$. 
d)$^{12}$CO(J=1-0) and NH$_{3}$(1,1) spectra towards Cir X-1 (l,b=322.116$^{\circ}$,0.033$^{\circ}$) and EGO G321.94-0.01, (l,b=321.923$^{\circ}$,0.013$^{\circ}$).
\label{f1}}
\end{figure}

Towards Cir X-1, three distinct $^{12}$CO(J=1-0) peaks $>$ 4$\sigma$ are seen in Nanten data (see Fig.~\ref{f1}d). The molecular mass, obtained from X-factor conversion between integrated CO intensity and H$_{2}$ column density, and average density of each peak within a 2.5$'$ radius circle encompassing the radio nebula towards Cir X-1 is  $\sim$10$^{4}$ M$_{\odot}$ (d/10kpc)$^{2}$ and  $\sim$100 cm$^{-3}$ (d/10kpc)$^{-1}$ respectively for each peak. The cloud at $-$75 km/s (see Fig.~\ref{f1}a) appears to flank Supernova remnant (SNR) G321.9-0.3 and is broad (~20 km/s FWHM), indicating the gas may be disrupted by the SNR which is at 5.5-6 kpc. The cloud at $-$30 km/s (see Fig.~\ref{f1}b) is part of a large gas feature, $\sim$0.3$^{\circ}$x0.8$^{\circ}$, towards the centre of which we observe emission from CS(J=1-0), H$_{2}$O and NH$_{3}$(1,1), indicating core densities $>$ 10$^{5}$ cm$^{-3}$; and coincides with Extended Green Object (EGO) and likely massive young stellar outflow region G321.94-0.01\cite{Cyganowski}. The cloud at 10 km/s (see Fig.~\ref{f1}c) overlapping Cir X-1 is narrow and localised to its vicinity. There is a possible faint (3$\sigma$) detection of NH$_{3}$(1,1) at this velocity (see Fig.~\ref{f1}d).

\begin{figure}[!htb]
\centerline{\psfig{file=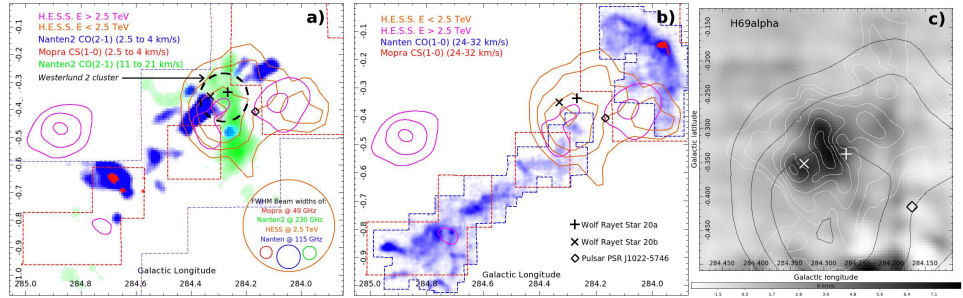,width=12.7cm}}
\vspace*{8pt}
\caption{Multiwavelength observations of the Westerlund 2 region. a) Velocity ranges chosen to show parent GMCs of Westerlund 2. a) \& b) Integrated Nanten2 $^{12}$CO(2-1) and Mopra CS(1-0) data\protect\cite{Furukawa} overlayed with smoothed H.E.S.S. significance contours for energies less than 2.5 TeV in increments of 1$\sigma$ from 4$\sigma$ and above 2.5 TeV in increments of 0.55$\sigma$ from 3$\sigma$.  See online version for colour, dashed lines indicate mapping regions. c) Mopra H69$\alpha$ RRL emission with low energy H.E.S.S contours (black) as in a) and MOST 843 MHz radio continuum contours (white).
\label{f2}}
\end{figure}

Two narrow, $\sim$3 km/s FWHM, peaks of CS(J=1-0) emission were seen towards Westerlund 2, indicating gas densities in those regions of $>$10$^{4}$ cm$^{-3}$. A peak of 0.31 K (5.8$\sigma$) at $\sim$3 km/s is coincident with a dense component of giant molecular cloud (GMC) located at 1.2-8.7 km/s and believed to be one of the two parent GMCs of Westerlund 2 (Fig.~\ref{f2}a). It is 2.7 pc in radius (d/5.4 kpc) and has a virial mass between 1.5x10$^{3}$ and 5.5x10$^{3}$ M⊙$_{\odot}$. The other peak of 0.26 K (7.5 $\sigma$) is seen in the arc, which coincides precisely with the peak of the CO emission and is at the same velocity range ($\sim$24-28 km/s) (see Fig.~\ref{f2}b). It is 2.8 pc in radius (d/7 kpc) and has a virial mass between 8.8x10$^{2}$ and 3.1x10$^{3}$ M$_{\odot}$. Virial masses are bounded by r$^{-2}$ and Gaussian density profile assumptions. Emission from radio recombination lines (RRLs) H62$\alpha$ (Fig.~\ref{f2}c), H65$\alpha$ and H69$\alpha$ was observed towards the HII region. It shows a morphological correspondence to the radio continuum emission and appears to be centred between the two Wolf-Rayet stars of the Westerlund 2 cluster. The emission in all three lines is broad ($\sim$45 km/s FWHM) and peaks at $\sim$0 km/s.

\section{Conclusions}
The ambiguity in associating peaks in the $^{12}$CO emission to Cir X-1 as well as the non-detection of any other molecular line emission has not enabled us to gain insights into the interaction between the relativistic jets and the ISM, nor to constrain the distance to Cir X-1. We tentatively associate the molecular gas located at 10 km/s with Cir X-1, and that at $-$75 km/s with SNR G321.9-0.3. EGO G321.94-0.01 has been associated with  $^{12}$CO(1-0), CS(1-0), H$_{2}$O and NH$_{3}$(1,1) emission at $-$30 km/s. Molecular cloud masses of the order of $\sim$10$^{4}$ M$_{\odot}$ towards Cir X-1 make $\gamma$-ray production likely given the jet energetics, and warrants deeper H.E.S.S observations.

Two peaks of CS are seen towards Westerlund 2, providing evidence of dense gas towards both the jet feature and a parent molecular cloud of the cluster, either of which may be associated with the high energy emission in the region. RRL emission highlights the extent of the partially ionised gas in the HII region RCW 49.

\section*{Acknowledgements}
The Mopra radio telescope is part of the Australia Telescope National Facility which is funded by the Commonwealth of Australia for operation as a National Facility managed by CSIRO. The Nanten project was based on a mutual agreement between Nagoya University and the Carnegie Institution of Washington. We acknowledge ARC grants DP0662810 and DP1096533.

\appendix


\begin{thebibliography}{00}  

\bibitem{Tudose} V. Tudose {\it et al.}, {\it MNRAS}, {\bf 390}, 447 (2008).

\bibitem{Stewart} R.T. Stewart {\it et al.}, {\it MNRAS}, {\bf 261}, 593 (1993).

\bibitem{Haynes} R.F. Haynes {\it et al.}, {\it Nature}, {\bf 324}, 233 (1986).

\bibitem{Sell} P. H. Sell {\it et al.}, {\it ApJ 719 L194}, {\bf 719}, 194 (2010).

\bibitem{Nicholas} H.E.S.S Collab. (B. Nicholas {\it et al}.), in {\it AIP Conf. Proc 1085}, eds. F.A. Aharonian, W. Hofmann and F. Rieger (Heidelberg, Germany, 2008), p.245.

\bibitem{Fukui09} Y. Fukui {\it et al.}, {\it PASJ}, {\bf 61}, 23 (2009).

\bibitem{Aharonian} H.E.S.S Collab. (A. Aharonian {\it et al}.), 
{\it A\&A} {\bf 476}, 1075 (2007).

\bibitem{Mizuno} A. Mizuno and Y. Fukui, in {\it ASP Conf. Ser. 317}, eds. D. Clemens, R. Shah and T. Brainerd (San Francisco, CA 2004), p.59

\bibitem{Maxted} N.I. Maxted {\it et al.}, {\it MNRAS}, {\bf 422}, 2230 (2012).

\bibitem{Cyganowski} C. J. Cyganowski {\it et al.}, {\it AJ}, {\bf 136}, 2391 (2008).

\bibitem{Bhattacharya} D. Bhattacharya and G. L. Case, {\it ApJ}, {\bf 504}, 761 (1998).

\bibitem{Furukawa} N. Furukawa {\it et al.}, submitted to {\it ApJ}

\end{thebibliography}
\end{document}